\documentclass[preprintnumbers,twocolumn,prl,amsmath,amssymb]{revtex4}

\usepackage{graphicx}
\usepackage{dcolumn}
\usepackage{bm}

\usepackage{hyperref}

\graphicspath{{Figures/}}

\begin{document}
\title{Efficient single photon emission from a high-purity hexagonal boron nitride crystal}
\author{L.~J.~Mart\'{\i}nez$^{1}$}
\author{T. Pelini$^{1}$}
\author{V. Waselowski$^{2}$}
\author{J. R. Maze$^{2}$}
\author{B. Gil$^{1}$}
\author{G. Cassabois$^{1}$}
\author{V. Jacques$^{1}$}
\email{vincent.jacques@umontpellier.fr}
\affiliation{$^{1}$Laboratoire Charles Coulomb, Universit\'{e} de Montpellier and CNRS, 34095 Montpellier, France}
\affiliation{$^{2}$Facultad de F\'{i}sica, Pontificia Universidad Cat\'{o}lica de Chile, Santiago 7820436, Chile}

\begin{abstract}
Among a variety of layered materials used as building blocks in van der Waals heterostructures, hexagonal boron nitride (hBN) appears as an ideal platform for hosting optically-active defects owing to its large bandgap ($\sim 6$~eV). Here we study the optical response of a high-purity hBN crystal under green laser illumination. By means of photon correlation measurements, we identify individual defects emitting a highly photostable fluorescence under ambient conditions. A detailed analysis of the photophysical properties reveals a high quantum efficiency of the radiative transition, leading to a single photon source with very high brightness. These results illustrate how the wide range of applications offered by hBN could be further extended to photonic-based quantum information science and metrology. 
\end{abstract}
\maketitle

Hexagonal boron nitride (hBN), also known as ``{\it white graphite}'', is a wide-bandgap material which is commonly used in the industry since several decades owing to its high thermal and chemical stability~\cite{Paine1990,Jiang2015}. Continuous improvement of growth techniques has recently enabled to decrease the density of intrinsic defects in the hBN matrix~\cite{Taniguchi2004,Taniguchi2007}. The resulting high-purity crystals have released the engineering potential of this material for the development of efficient light-emitting devices in the deep ultraviolet~\cite{Kubota2007,Taniguchi2009}. More recently, hBN has even attracted a renewed interest because it exhibits a two-dimensional (2D) honeycomb structure similar to graphene, which can be used as a building block for the design of complex van der Waals heterostructures~\cite{Dean2010,Geim2013,Cui2015,Withers2015}. Given this wide range of promising applications, it is essential to precisely characterize intrinsic defects hosted in the hBN crystal in order to further optimize the properties of this material. In addition, the possibility to isolate defects at the ``{\it single-atom}'' level would offer new opportunities for applications of 2D materials in quantum technologies~\cite{Awschalom2013,Awschalom2015}.\\
\indent In early experiments, the characterization of point-defects and stacking faults was mainly realized through ensemble measurements using cathodoluminescence and optical spectroscopy in the deep ultraviolet~\cite{Lauret,Museur,Bourrellier,Cassabois2016a}. Making use of hBN/graphene heterostructures, it was recently shown that impurities hosted in the hBN matrix can also be detected by measuring local modifications of the conductivity in the graphene layer~\cite{Ju2014,Wong2015}. Using scanning tunnelling microscopy, this original approach was even used to isolate and manipulate {\it individual} defects with nanoscale resolution~\cite{Wong2015}.\\
 \begin{figure}[t]
\includegraphics[width = 8.9cm]{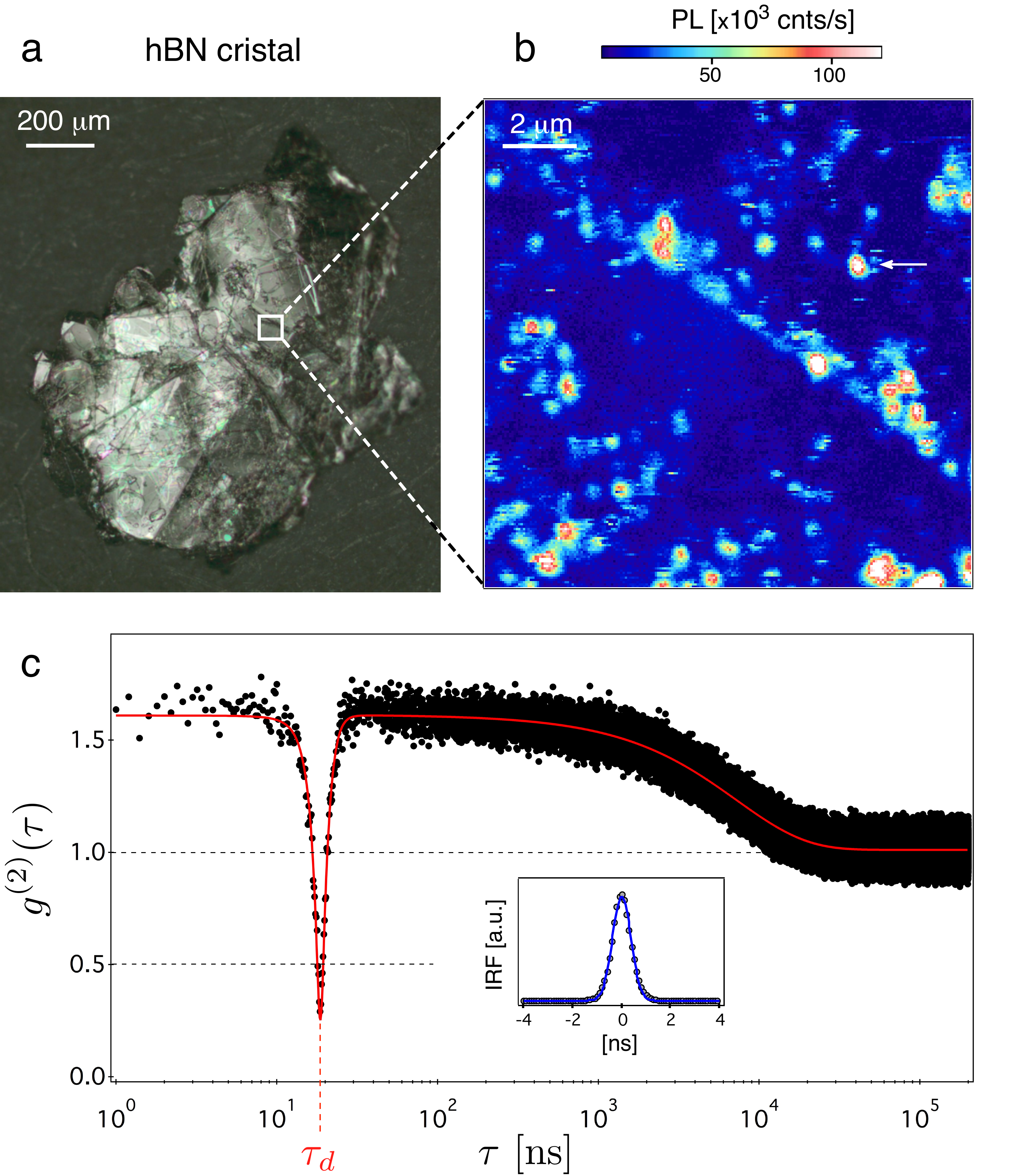}
\caption{(a) Microscope image of a high-purity hBN crystal from HQ graphene. (b) PL raster scan of the sample. (c) $g^{(2)}(\tau)$ function recorded from the PL spot shown by the white arrow in (b) for a laser excitation power $\mathcal{P}=150 \ \mu$W and a bin width $w=200$~ps. The solid line is data fitting using a three-level model while including the IRF of the detection setup [see main text]. For this experiment we obtain $\lambda_1=0.51$~ns$^{-1}$, $\lambda_2=0.14$~$\mu$s$^{-1}$ and $a=0.6$. Inset: IRF measured using 50-ps laser pulses. The solid line is a Gaussian fit. 
}
\label{Fig1}
\end{figure}
\indent In this Letter, we follow a conceptually simpler strategy to probe individual defects in a high-purity hBN crystal by measuring the photoluminescence (PL) response under optical illumination with an energy much lower than the bandgap~\cite{Tran_NatNano2016,Tran_arXiv2016,Fuchs2016,Tran_PRAp2016}. This method is relatively easy to implement in hBN owing to its very large bandgap $E_g\sim 6$~eV~\cite{Cassabois2016}. Here we isolate individual defects emitting around $2$~eV with a high photostability under ambient conditions. A detailed analysis of the photophysical properties indicates a highly efficient radiative transition leading to one of the brightest single photon source reported to date. \\ 
\indent The optical response of a commercial high-purity hBN crystal [Fig.~1a] was analyzed using a scanning confocal microscope operating under ambient conditions. Optical illumination was carried out at the wavelength $\lambda=532$~nm, which allows to probe deep defects with energy levels far within the bandgap of hBN. The excitation laser was focused onto the sample with a high numerical aperture oil-immersion microscope objective (NA=1.35) mounted on xyz-piezoelectric scanner. The PL response of the hBN crystal was collected by the same objective and spectrally filtered from the remaining excitation laser with a razor-edge longpass filter at $532$~nm (Semrock). The collected PL was then focused in a 50-$\mu$m-diameter pinhole and finally directed either to a spectrometer or to silicon avalanche photodiodes operating in the single-photon counting regime. A typical PL raster scan of the sample is shown in Fig.~1b, revealing bright diffraction-limited spots, which correspond to the emission from deep defects hosted in the hBN matrix. \\  
\indent The single-atom nature of the emitters was verified by measuring the second-order correlation function $g^{(2)}(\tau)$ using two avalanche photodiodes, D$_1$ and D$_2$, mounted in a Hanbury-Brown and Twiss (HBT) configuration. Such a detection scheme is commonly used in a ``{\it start-stop}'' mode to record the histogram of time intervals $K(\tau)$ between two {\it consecutive} single-photon detections. Once properly normalized to a Poissonian light source, the recorded histogram is equivalent to the second-order correlation function $g^{(2)}(\tau)$ provided that $\tau\ll R^{-1}$, where $R$ is the photon detection rate~\cite{Reynaud1983,Liu2016}. We stress that significant deviations from this equivalence have been experimentally evidenced as soon as $\tau\gtrsim100$~ns for $R\sim 10^5$~s$^{-1}$~\cite{Fleury2000}. The ``{\it start-stop}'' method is thus limited to correlation measurements at very short time scales. \\ 
 \begin{figure}[b]
\includegraphics[width = 8.3cm]{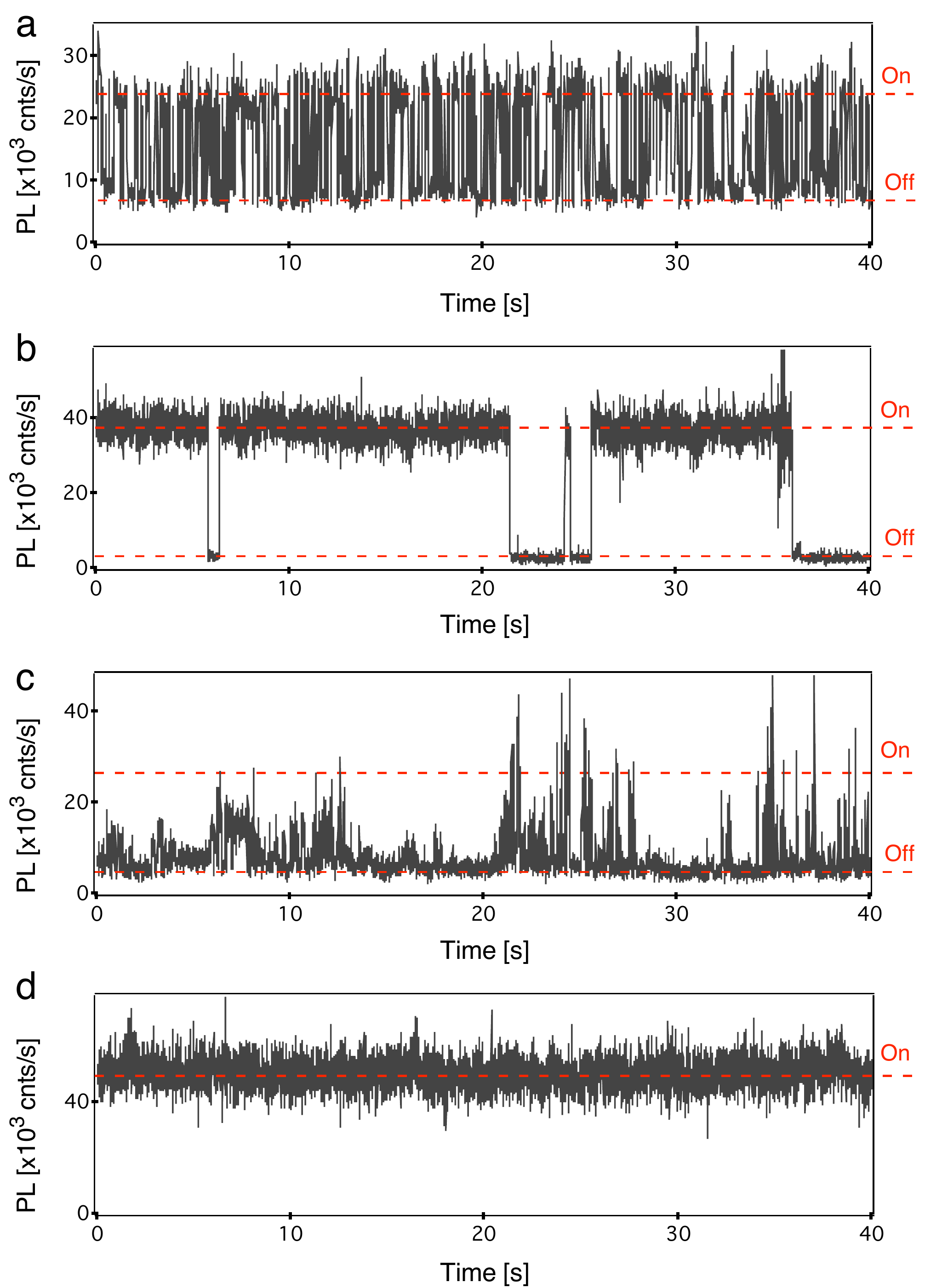}
\caption{(a) to (d) Typical PL time traces recorded from different individual defects in a high purity hBN cristal. The laser excitation power is set to $\mathcal{P}=10 \ \mu$W. Bin size: $10$~ms.}
\label{Fig1}
\end{figure}
\indent In order to record the $g^{(2)}(\tau)$ function without any temporal restrictions, the HBT setup was rather used to measure $J(\tau)$, defined as the number of photons detected at time $\tau$ provided that a photon is detected at time $t=0$. To this end, a photon detection event on detector D$_1$ was used to trigger the acquisition of a PL time trace on detector D$_2$ using a large-range time-to-digital converter (FastComtec, P7889). After $\mathcal{N}$ repetitions of the measurement, the resulting time trace $J(\tau)$ is directly linked to the second-order correlation function through $g^{(2)}(\tau)=J(\tau)/(\mathcal{N}wR_2)$, where $w$ is the bin time and $R_2$ is the photon detection rate on detector D$_{2}$. Experimentally, an electronic delay $\tau_d$ was introduced in order to obtain the full profile of the $g^{(2)}(\tau)$ function at short time scale.\\ 
\indent Figure~1c shows a typical $g^{(2)}(\tau)$ measurement recorded over five temporal decades from an isolated PL spot of the hBN cristal. The anticorrelation effect observed at short time scale, $g^{2}(\tau_d)\approx 0.25<0.5$, is the signature of single photon emission from an individual defect. The deviation from an ideal anticorrelation [$g^{2}(\tau_d)=0$] cannot be explained by a residual background PL from the host matrix owing to the high value of the signal-to-background ratio ($>20$)~\cite{Beveratos_EPJD2002}. It rather results from the instrumental response function (IRF) of the detection setup, which is mainly fixed by the time jitter of the single photon detectors [see inset in Fig.~1c]~\cite{Wu2006}. The correlation function also reveals photon bunching $g^{(2)}(\tau)>1$, which indicates that optical cycles involve a non-radiative relaxation path through a long-lived metastable level. Modeling the defect as a three-level system [see Fig.~4a], the second order correlation function can be expressed as~\cite{Rarity1998}
\begin{equation}
 g^{(2)}(\tau)=1-(1+a)e^{-\lambda_1|\tau-\tau_d|}+ae^{-\lambda_2|\tau-\tau_d|} \ ,
 \label{G2short}
 \end{equation}
where $a$ is the photon bunching amplitude and $\lambda_1$ (resp. $\lambda_2$) is the decay rate of the anticorrelation (resp. bunching) effect~\cite{Note}. As shown in Fig.~1c, the experimental data are well fitted by the convolution of Eq.~(\ref{G2short}) with the independently measured IRF of the detection setup. A detailed analysis of the dynamics of optical cycles by means of photon correlation measurements is given in the last section of the paper.

 \begin{figure*}[t]
\includegraphics[width=14.5cm]{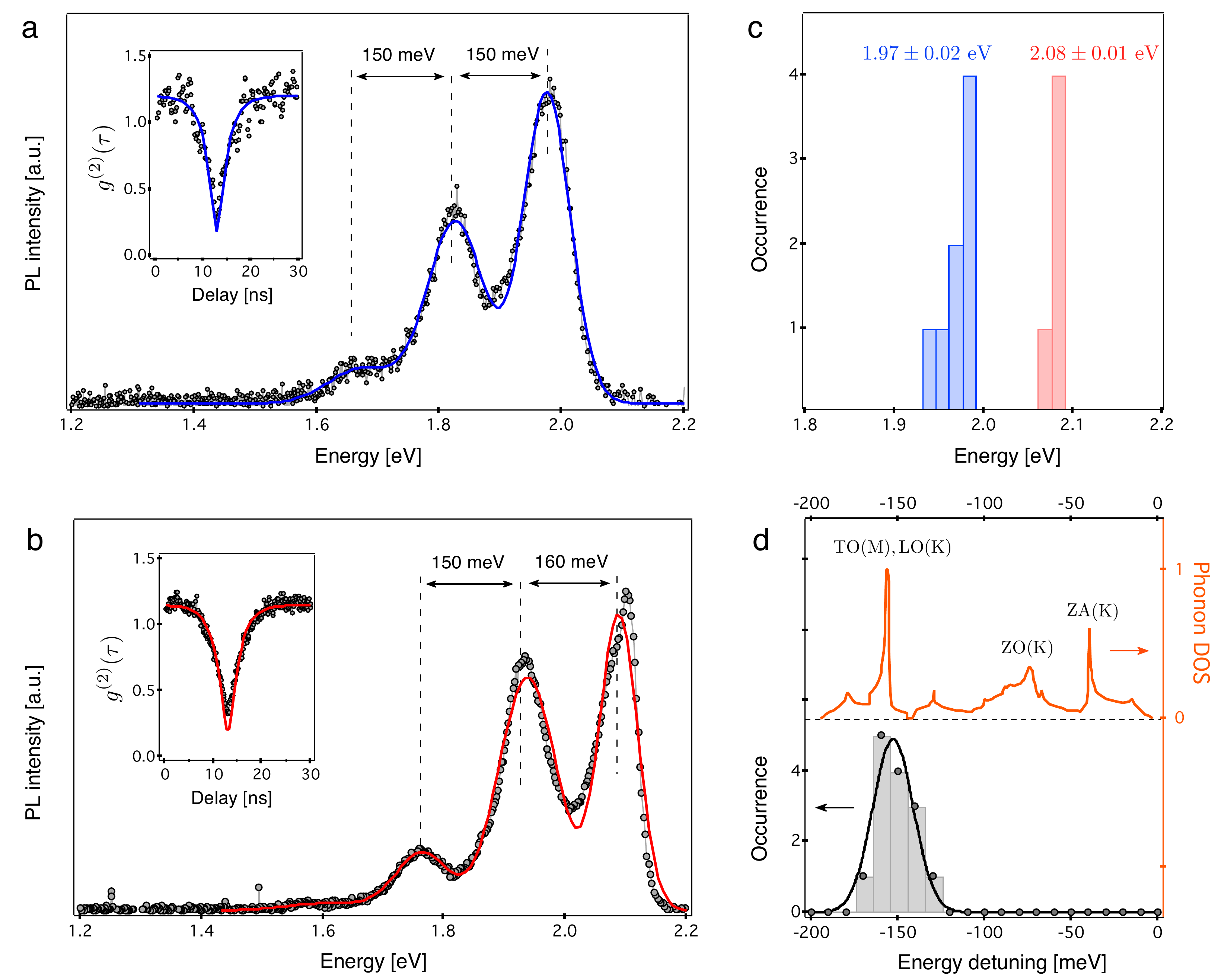}
\caption{(a,b) PL spectra recorded from individual defects in hBN at room temperature under green laser illumination. The solid lines are data fitting with Gaussian functions. The insets show the corresponding second-order correlation function $g^{(2)}(\tau)$. (c) Histogram of the ZPL energy for a set of 13 single defects. Two families of defects can be observed with ZPL energies at $1.97\pm0.02$~eV and $2.08\pm0.01$~eV. (d) Bottom: Histogram of the energy detuning between the ZPL and the phonon replica. Top: phonon density of states (DOS) in hBN extracted from Ref.~\cite{Kern_PRB1999}. 
}
\label{Fig2}
\end{figure*}
 An important property of any individual defect acting as a single photon source is its photostability over time. Representative PL time traces recorded at low laser power from individual defects hosted in hBN are shown in Fig.~2. Most defects exhibit a pronounced blinking behavior, as often observed for solid-state emitters at room temperature like quantum dots~\cite{Nirmal1996}, dye molecules~\cite{Dickson1997} and deep defects in wide-bandgap materials~\cite{Zn0_2012,Zn0_Igor2014,Bradac2010,Castelletto2014}. The switching rate between the {\it on} and {\it off} states varies significantly from one emitter to another [Fig. 2a-c], which suggests that the blinking dynamics are strongly linked to the local environment of the defect. Such a blinking could originate from optically-induced charge state conversion mediated by a charge exchange with other defects acting as electron donors/acceptors. Permanent photobleaching, {\it i.e.} the irreversible conversion of an optically-active defect into a non-fluorescent entity, was also observed after few minutes of optical illumination for most emitters. However, we stress that around $5\%$ of the emitters exhibit a {\it perfect photostability} over time [Fig.~2d], even under high laser excitation power, as observed for few types of defects in diamond and SiC at room temperature~\cite{Igor2011,Englund2016}. This ratio might be improved through annealing procedures and/or chemical modifications of the sample surface~\cite{Bradac2010,Castelletto2014}. On average, one stable defect could be found in each $50\times50 \ \mu$m$^2$ scan of the sample.
 
  \begin{figure*}[t]
\includegraphics[width = 18cm]{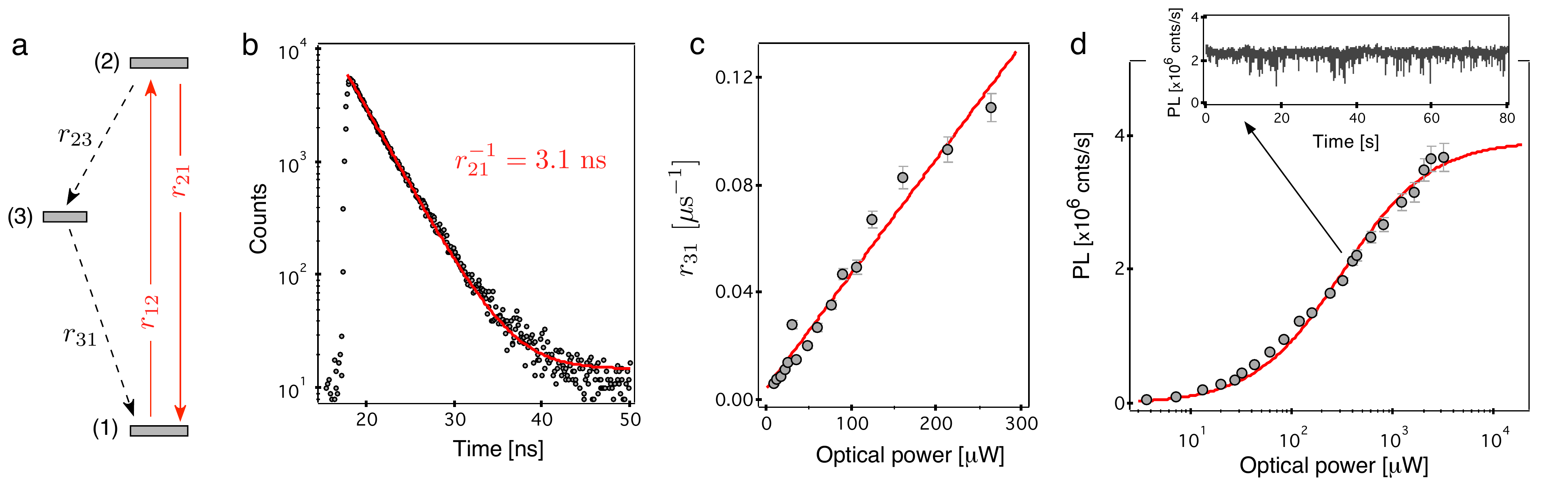}
\caption{(a) Three-level model including a radiative transition between the ground (1) and excited (2) levels, and a non-radiative decay through a metastable level (3). The coefficient $r_{ij}$ denotes the transition rate from level $(i)$ to level $(j)$. (b) PL decay recorded under optical excitation with 50-ps laser pulses. The solid line is data fitting with a single exponential function. (c)~Decay rate of the metastable state $r_{31}$ as a function of the laser excitation power. The solid line is data fitting with a linear function [see main text]. (d) PL saturation curve of the defect. The inset shows a PL time trace recorded for a laser excitation power $\mathcal{P}=400 \ \mu$W. All these measurements are performed on the same individual defect.}
\label{Fig}
\end{figure*}
Typical emission spectra recorded at room temperature from individual emitters are shown in Figs.~3(a,b). We identify two families of spectra with zero-phonon line (ZPL) energies at $1.97\pm0.02$~eV ($629$~nm) and $2.08\pm0.01$~eV ($596$~nm) [Fig.~3(c)]. The overall structure of the spectra is very similar for the two families, including well-resolved phonon replica with a characteristic energy detuning of $150\pm 10$~meV [Fig.~3(d)]. A similar energy spacing was recently reported in the case of single defects in multilayer hBN flakes~\cite{Tran_NatNano2016,Tran_arXiv2016,Fuchs2016}. In fact, this value fairly matches an extremum of the phonon density of states in bulk hBN linked to transverse (TO) and longitudinal (LO) optical phonons at the K and M points of the Brillouin zone [Fig.~3(d)]~\cite{Kern_PRB1999}. Our defects being linked to deep levels with a transition energy much smaller than the $6$~eV-bandgap of hBN, we expect an extension of the electronic wavefunction comparable to, or smaller than the lattice parameter of the hBN matrix~\cite{Jaros}. As a consequence, the defect wavefunction in $k$-space is spread over the whole Brillouin zone with a significant coupling to the zone-edge phonons, {\it e.g.}~at the K and M points. Since the interaction with optical phonons is much more efficient than with acoustic ones~\cite{Cassabois2016}, we interpret the systematic $150$~meV-energy detuning of the phonon replica as resulting from a dominant coupling to the zone-edge optical phonons of highest density of states [see Fig.~3(d)].

Individual defects in wide-bandgap materials can often be found in various charge states having very different optical properties. A well-known example is the nitrogen-vacancy (NV) defect in diamond, whose ZPL energy is shifted by $\approx 200$~meV depending of the charge state NV$^{0}$/NV${^-}$ of the defect~\cite{Rondin2009,Hauf2011}, and becomes even optically inactive in its positively-charged state NV$^{+}$~\cite{Grotz2012,Meijer2015}. Using scanning tunnelling microscopy (STM), it was recently shown that individual defects in hBN can also be found in positive, negative and neutral charge states. Active charge state manipulation was even demonstrated by tuning the Fermi energy level with respect to the defect's charge transition level by locally applying an electric field with a STM tip~\cite{Wong2015}. We tentatively attribute the two groups of spectra shown in Fig.~3 to different charge states of the same defect, while the small variation of the ZPL energy in each group ($\pm20$~meV) might result from local variations of the strain in the hBN matrix. Optically-induced charge state conversion towards a dark-state might also play an important role in the blinking/bleaching dynamics of the defect.

 \indent We now investigate the dynamics of optical cycles for a stable defect modeled as a three-level system [Fig.~4a]. The lifetime of the excited level was first measured through time-resolved PL measurements under pulsed laser excitation [Fig.~4b]. The PL decay is well fitted by a single exponential function, leading to a lifetime of the excited level $r_{21}^{-1}=3.1\pm0.1$~ns. This value is in good agreement with the one recently measured in hBN flakes for individual defects showing similar spectral properties~\cite{Tran_NatNano2016,Tran_arXiv2016}. To gain insights into the dynamics of the metastable level, the characteristic parameters of the bunching effect ($a,\lambda_2$) were inferred from $g^{(2)}(\tau)$ measurements recorded at different laser excitation power $\mathcal{P}$ [see Eq.~(1)]. Using a three-level model, the relaxation rate of the metastable level $r_{31}$ is then simply given by~\cite{Rarity1998} 
 \begin{equation}
 r_{31}=\frac{\lambda_2}{1+a} \ .
\end{equation}
As shown in Fig.~4c, our results indicate a pronounced increase in $r_{31}$ with the laser power, which suggests that optical illumination induces deshelving of the metastable level [Fig.~2c]. This effect is often observed for individual emitters in solid-state systems~\cite{Fleury2000,Wu2006}. By fitting the data with a linear power-dependence of the deshelving process $r_{31}=r_{31}^0(1+\beta \mathcal{P})$, we obtain an estimate of the metastable level lifetime $[r_{31}^0]^{-1}=210\pm 80 \ \mu$s.

The PL rate $R$ was finally measured as a function of $\mathcal{P}$ [Fig.~4d]. Data fitting with a simple saturation function $R=R_{\infty}/(1+\mathcal{P}/\mathcal{P}_{\rm sat})$ leads to a saturation power $\mathcal{P}_{\rm sat}=310\pm20 \ \mu$W and an emission rate at saturation $R_{\infty}=3.9\pm0.1\times10^{6}$ counts.s$^{-1}$, corresponding to one of the brightest single photon source reported to date. For comparison, the emission rate at saturation is $\sim 20$ times larger than the one obtained from a single nitrogen-vacancy (NV) defect in diamond with the same confocal microscope. Such high counting rates indicate that the short excited-level lifetime results from a strong radiative oscillator strength with near-unit quantum efficiency rather than PL quenching involving fast non-radiative decay processes ($r_{23}\ll r_{21}$). \\

To summarize, we have isolated individual defects in a commercial high-purity hBN crystal, leading to a robust single photon source with high brightness under ambient conditions. The analysis of the photophysical properties of the defect provides important informations for understanding its structure using {\it ab-initio} calculations~\cite{Attaccalite2011}. These results pave the road towards applications of hBN, and more generally van der Waals heterostructures, in photonic-based quantum information science~\cite{OBrien2009} and optoelectronics~\cite{Ju2014}. \\

\noindent {\it Acknowledgements:} We thank M. Kociak, A. Zobelli, J.-F. Roch and I. Robert-Philip for fruitful discussions. This work was financially supported by the network GaNeX (ANR-11-LABX-0014). GaNeX belongs to the publicly funded {\it Investissement d'Avenir} program managed by the French ANR agency. The authors acknowledge support from Ecole Normale Sup\'erieure de Cachan, Conicyt-Fondecyt (grant No.1141185), Conicyt-PCHA Doctoral program (grant No 2013-21130747) and PMI PUC 1203. G.C. is member of Institut Universitaire de France.


\begin{thebibliography}{30}

\bibitem{Paine1990}
R. T. Paine and C. K. Narula, {\it Chem. Rev.} {\bf 90}, 73-91 (1990).

\bibitem{Jiang2015}
X. F. Jiang, Q. H. Weng, X. B. Wang, X. Li, J. Zhang, D. Golberg and Y. Bando, {\it J. Mater. Sci. Technol.} {\bf 31}, 589-598 (2015).

\bibitem{Taniguchi2004}
K. Watanabe, T. Taniguchi, and H. Kanda, {\it Nat. Mater.} {\bf 3}, 404-409 (2004).

\bibitem{Taniguchi2007}
T. Taniguchi and K. Watanabe, {\it J. Cryst. Growth} {\bf 303}, 525-529 (2007).

\bibitem{Kubota2007}
Y. Kubota, K. Watanabe, O. Tsuda, and T. Taniguchi, {\it Science} {\bf 317}, 932-934 (2007).

\bibitem{Taniguchi2009}
K. Watanabe, T. Taniguchi, T. Niiyama, K. Miya, and M. Taniguchi, {\it Nat. Phot.} {\bf 3}, 591-594 (2009). 

\bibitem{Dean2010}
C. R. Dean, {\it et al.}, {\it Nat. Nano.} {\bf 5}, 722-726 (2010).

\bibitem{Geim2013}
A. K. Geim and I. V. Grigorieva, {\it Nature} {\bf 499}, 419-425 (2013).

\bibitem{Cui2015}
X. Cui, {\it et al.}, {\it Nat. Nano.} {\bf 10}, 534Ð540 (2015). 

\bibitem{Withers2015}
F. Withers, {\it et al.}, {\it Nat. Mater.} {\bf 14}, 301Ð306 (2015). 

\bibitem{Awschalom2013}
D. D. Awschalom, L. C. Bassett, A. S. Dzurak, E. L. Hu, and J. R. Petta, {\it Science} {\bf 339}, 1174-1179 (2013).

\bibitem{Awschalom2015}
W. F. Koehla, H. Seoa, G. Gallia, and D. D. Awschalom, {\it MRS Bulletin} {\bf 40}, 1146 (2015).

\bibitem{Lauret}
M. G. Silly, P. Jaffrennou, J. Barjon, J.-S. Lauret, F. Ducastelle, A. Loiseau, E. Obraztsova, B. Attal-Tretout, and E. Rosencher, {\it Phys. Rev. B} {\bf 75}, 085205 (2007).

\bibitem{Museur}
L. Museur, E. Feldbach, and A. Kanaev, {\it Phys. Rev. B} {\bf 78}, 155204 (2008).

\bibitem{Bourrellier}
R. Bourrellier {\it et al.}, {\it ACS Phot.} {\bf 1}, 857 (2014)

\bibitem{Cassabois2016a}
G. Cassabois, P. Valvin and B. Gil, {\it Pys. Rev. B} {\bf 93}, 035207 (2016).

\bibitem{Ju2014}
L. Ju, {\it et al.}, {\it Nat. Nano.} {\bf 9}, 348-352 (2014). 

\bibitem{Wong2015}
D. Wong {\it et al.}, {\it Nat. Nano.} {\bf 10}, 949-954 (2015).

\bibitem{Tran_NatNano2016}
T. T. Tran, K. Bray, M. J. Ford, M. Toth, and I. Aharonovich, {\it Nat. Nano.} {\bf 11}, 37-41 (2016).

\bibitem{Tran_arXiv2016}
T. T. Tran, C. El Badawi, D. Totonjian, C. J. Lobo, G. Grosso, H. Moon, D. R. Englund, M. J. Ford, I. Aharonovich, and M. Toth, preprint arXiv:1603.09608 (2016).

\bibitem{Fuchs2016}
N. R. Jungwirth, B. Calderon, Y. Ji, M. G. Spencer, M. E. Flatt\'e, and G. D. Fuchs, preprint arXiv:1605.04445 (2016).

\bibitem{Tran_PRAp2016}
T. T. Tran, {\it et al.}, {\it Phys. Rev. Applied} {\bf 5}, 034005 (2016). 

\bibitem{Cassabois2016}
G. Cassabois, P. Valvin, and B. Gil, {\it Nat. Phot.} {\bf 10}, 262-266 (2016).

\bibitem{Reynaud1983}
S. Reynaud, {\it Ann. Phys. Fr.} {\bf 8}, 315-370 (1983).

\bibitem{Liu2016}
C.-H. Huang, Y.-H. Wen, and Y.-W. Liu, {Opt. Express} {\bf 24}, 4278-4288 (2016).

\bibitem{Fleury2000}
L. Fleury, J.-M. Segura, G. Zumofen, B. Hecht, and U. P. Wild, {\it Phys. Rev. Lett.} {\bf 84}, 1148-1151 (2000).

\bibitem{Beveratos_EPJD2002}
A. Beveratos A, S. K$\ddot{\rm u}$hn, R. Brouri, T. Gacoin, J. P. Poizat, and P. Grangier, {\it Eur. Phys. J D} {\bf 18}, 191-196 (2002).

\bibitem{Wu2006}
E Wu, V. Jacques, H. Zeng, P. Grangier, F. Treussart, and J.-F. Roch, {\it Opt. Express} {\bf 14}, 1296-1303 (2006).
 
 \bibitem{Rarity1998}
S. C. Kitson, P. Jonsson, J. G. Rarity, and P. R. Tapster, {\it Phys. Rev. A} {\bf 58}, 620-627 (1998).

\bibitem{Note}
This expression is valid if $(r_{23},r_{31})\ll r_{21}$.

\bibitem{Nirmal1996}
M. Nirmal, B. O. Dabbousi, M. G. Bawendi, J. J. Macklin, J. K. Trautman, T. D. Harris, and L. E. Brus, {\it Nature} {\bf 383}, 802-804 (1996).

\bibitem{Dickson1997}
R. M. Dickson, A. B. Cubitt, R. Y. Tsien, and W. E. Moerner, {\it Nature} {\bf 388}, 355-358 (1997).

\bibitem{Zn0_2012}
S. Choi, B. C. Johnson, S. Castelletto, C. Ton-That, M. R. Phillips, and I. Aharonovich, {\it Appl. Phys. Lett.} {\bf 104}, 261101 (2014)

\bibitem{Zn0_Igor2014}
A. J. Morfa, B. C. Gibson, M. Karg, T. J. Karle, A. D. Greentree, P. Mulvaney, and S. Tomljenovic-Hanic, {\it Nano Lett.} {\bf 12}, 949-954 (2012).

\bibitem{Bradac2010}
C. Bradac, {\it et al.}, {\it Nat. Nano.} {\bf 5}, 345-349 (2010)

\bibitem{Castelletto2014}
S. Castelletto, B. C. Johnson, V. Iv\'ady, N. Stavrias, T. Umeda, A. Gali, and T. Ohshima, {\it Nat. Mater.} {\bf 13}, 151-156 (2014).

\bibitem{Igor2011}
I. Aharonovich, S. Castelletto, D. A. Simpson, C.-H. Su, A. D. Greentree, and S. Prawer, {\it Rep. Prog. Phys.} {\bf 74}, 076501 (2011).
 
\bibitem{Englund2016}
B. Lienhard, T. Schr$\ddot{\rm o}$der, S. Mouradian, F. Dolde, T. T. Tran, I. Aharonovich, and D. R. Englund, preprint arXiv:1603.05759 (2016).

\bibitem{Kern_PRB1999}
G. Kern, G. Kresse, and J. Hafner, {\it Phys. Rev. B} {\bf 59}, 8551 (1999).
    
\bibitem{Jaros}
M. Jaros and S. F. Ross, {\it J. Phys. C: Solid St. Phys.} {\bf 6}, 1753 (1973).

\bibitem{Rondin2009}
L. Rondin {\it et al.}, {\it Phys. Rev. B} {\bf 82}, 115449 (2010).

\bibitem{Hauf2011}
M. V. Hauf {\it et al.}, {\it Phys. Rev. B} {\bf 83}, 081304 (2011).

\bibitem{Grotz2012}
B. Grotz {\it et al.}, {\it Nat. Commun.} {\bf 3}, 729 (2012).

\bibitem{Meijer2015}
C. Schreyvogel, V. Polyakov, R. Wunderlich, J. Meijer, and C. E. Nebel, {\it Sci. Rep.} {\bf 5}, 12160 (2015).

\bibitem{Attaccalite2011}
C. Attaccalite, M. Bockstedte, A. Marini, A. Rubio, and L. Wirtz, {\it Phys. Rev. B} {\bf 83}, 144115 (2011).

\bibitem{OBrien2009}
J. L. O'Brien, A. Furusawa, and J. Vuckovi\'c, {\it Nat. Phot.} {\bf 3}, 687-695 (2009). 

\end{thebibliography}
\end{document}